# Towards Blockchain-Based Secure Data Management for Remote Patient Monitoring

Md Jobair Hossain Faruk*, Hossain Shahriar†, Maria Valero†, Sweta Sneha‡,
Sheikh I. Ahamed§ and Mohammad Rahman¶

*Software Engineering Department, Kennesaw State University
†Information Technology Department, Kennesaw State University
‡Information Systems Department, Kennesaw State University
§Computer Science Department, Marquette University
¶ Electrical and Computer Engineering Department, Florida International University

*Abstract*—Traditional data collection, storage and processing of Electronic Health Records (EHR) utilize centralized techniques that pose several risks of single point of failure and lean the systems to a number of internal and external data breaches that compromise their reliability and availability. Blockchain is an emerging distributed technology that can solve these issues due to its immutability and architectural nature that prevent records manipulation or alterations. In this paper, we discuss the progress and opportunities of remote patient monitoring using futuristic blockchain technologies and its two primary frameworks: Ethereum and Hyperledger Fabric. We also discuss the possible blockchain use cases in software engineering for systematic, disciplined, and quantifiable application development. The study extends by introducing a system architecture for EHR data management using Ethereum as a model. We discuss the challenges and limitations along with the initial evaluation results of the proposed system and draw future research directions in this promising area.

## I. INTRODUCTION

Health can be defined as "a state of physical, mental, and social prosperity and not merely the absence of disease or infirmity." Having proper healthcare is a demand of human beings in the modern world [1]. Good quality of care and better access to healthcare facilities are paramount importance for society and elderly population. Statistics indicate that from 1980 to 2016, the average life expectancy at birth increased from 73.7 to 78.6 years [2]. As a result, the healthcare industry is an important sector that seems to emerge as one of the essential part of human lives. The current scene of computerized wellbeing is not only technological, but also social, cognitive, and political; the ultimate goal is to cooperate with participatory health, a partnership with digital devices collecting data and generating insights, adopt new models of care, evolving through collaborations between clinicians, patients, and carers [3].

The term of health informatics is the study and implementation of methods to enhance the administration of patient information, clinical knowledge, demographic data, and other information related to patient care and community health [4]. Health informatics encompasses not just the use of computers, but also the complete management of information in healthcare, including the creation and evaluation of methods and systems for acquiring, processing, and interpreting patient data [5]. According to a study by Hassan Aziz [6], health informatics is a wide-ranging science that encompasses the complex mixture of people, organizations, illnesses, patient care, and treatment all of which are intertwined with modern information technology, particularly in computing and communication.

Health records used to be recorded on paper, kept in folders divided into categories based on the type of note, and there was only one copy available. Such changes appeared in the 1960s and 1970s when new computer technologies were developed to support Electronic Medical Record (EMR) [7]. EMRs, grew in popularity because of it's ability to rapidly gather and manage sets of information, monitor changes in patient outcomes after installation of a new practice or treatment, and determine whether patients are due for physical exams, procedures, immunizations [8]. Correspondingly, EHRs have revolutionized the format of health records, transforming the health-care industry and making patients' medical records easier to read and access from virtually anywhere on the globe [7]. However, all the records in the conventional computing approach can be manipulated or altered easily, which creates concerns in terms of security and privacy of patients [9]. This is where an approach called "blockchain" emerged to introduce a revolutionary computer protocol used for the digital recording and storing of information in a decentralized and distributed ledger [10, 11].

A solution was crucial because data in existing telehealth and telemedicine systems are stored and processed centrally, which increases the danger of a single point of failure and exposes the systems to a number of external and internal threats [12]. However, data should never be stored as writing in pencil that could deface anytime rather storing in an immutable format that will protect data transaction transparently, emphasizes by the author H. Jobair. In order to reform the traditional healthcare practices, blockchain technology can be a model that helps to address such crucial problems [12, 13]. Blockchain is potentially a solution due to its immutability and architectural nature, where every block has a particular summary of the preceding block that is arranged using a secure hash value, string order, timing, content, and order of trades that can not be manipulated or altered. [14]. The key contributions of this study are listed below:

- We discuss existing progress and the prospective opportunities for Remote Patient Monitoring using Blockchain technology.
- We provide a thorough overview of blockchain technology and its two widely frameworks, (i) Ethereum and (ii) Hyperledger Fabric.
- We contribute by imparting an adequate overview of the concept of blockchain technology in software engineering and its inter alia interaction.
- We propose a system architecture for data management of remote patient monitoring using Ethereum as a model

The remainder of the paper is organized as follows: we present a synopsis of blockchain technology and its associated frameworks in Section II followed by reviewing the relevant literature and proposed systems in Section III. Section IV presents an ethereum-based prototype for remote patient monitoring. We draw a discussion on architectural challenges and limitations in Section V. Finally, Section VI provides some concluding comments and future research directions. 1mm vertical space

## II. BLOCKCHAIN TECHNOLOGY

The application of blockchain in healthcare is a recent addition in trusted sharing of sensitive healthcare information. It is defined as a distributed, incorruptible database of records or digital events which is executed, validated, and maintained by a network of computers instead of a single central network among participating parties around the world [15, 16]. Blockchain, according to the Organization for Economic Co-operation and Development (OECD) [17], is a combination of currently existing technologies that can be used to establish networks using distributed ledger technology (DLT). These networks can store data between a group of users that are authenticated by cryptographic tools and agreed upon through predefined network protocols, usually without the control of a central authority. The concept of blockchain is completely opposite to traditional methods; while the conventional approach stores data in a centralized database, blockchain stores data in a decentralized way. Blockchain records a timestamp to avoid tempering the stored data. This novel approach was first devised to run Bitcoin cryptocurrency, but it is now being advocated by different industries including healthcare due to its enhanced authentication, confidentiality, transparency, and unique data sharing characteristics verified by consensus. In Fig. 1, proposed by Seyednima Khezr et al. [18] depicts a workflow of blockchain-based healthcare applications.

Blockchain technology allows the creation of a tamper-evident, shared, and trusted ledger that sequentially appends cryptographically secure data transactions. The ledger would only be accessible to trusted parties. The cryptographic techniques used to record information to a blockchain guarantee that once a transaction has been added to the ledger, it cannot be modified; thereby assuring participants that they are working with data transactions that are up-to-date, accurate, and nearly impossible to manipulate. The blockchain thus functions as the sole source of truth.

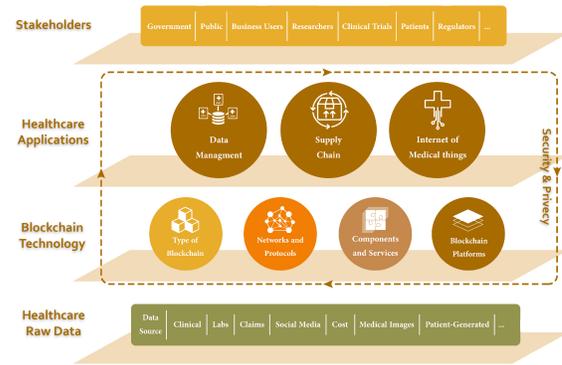

Fig. 1. represents blockchain-based workflow consists of primary layers including raw data, and stakeholders [18]

By narrowing the focus of the research to immutable data storing, implementation of telehealth and 21st-century patient data management, blockchain technology is crucial. And the immutable nature of said technology could lead to reduced cost of regulatory compliance with greater transparency, improved traceability, increased speed, and efficiency.

Multiple blockchain infrastructures have emerged, (i) permissionless blockchain that focuses on "trustless" networks used by any individuals. Bitcoin is an example of this kind since it is wide-open, permissionless, and anyone can buy bitcoins. (ii) permissioned blockchains where only pre-verified users shall have access which is vital for some enterprise-based systems in order to protect the business affairs [19, 20]. Ethereum and Hyperledger Fabric framework are two widely known blockchain-based approaches, where Hyperledger is a fully permissioned network designed for operations involving sensitive and confidential data, whereas Ethereum is a public network that enables permissioned networks [21].

### A. Ethereum

Ethereum was introduced as a platform in 2013 as a project which attempts to build a generalized blockchain technology in where all transaction-based state machine concepts can be built, with the goal of providing a tightly integrated end-to-end system to the end-developer. This allows building software on a hitherto unexplored compute paradigm in the mainstream: a trustful object messaging compute framework [22]. Ethereum represents a blockchain with built-in decentralized transactions and a turing-complete execution environment where the system can perform any computations. However, all nodes must have access to the whole records in blockchain. A Merkle Patricia Tree (MPT) is being used to improve the state. MPT is a special type of data structure that may store cryptographically authenticated data as keys and values [23]. Fig. 2 depicts the block structure of Ethereum and its Merkle Tree in which the hash of the root node (the tree's initial node) is dependent on the hashes of all sub-nodes.

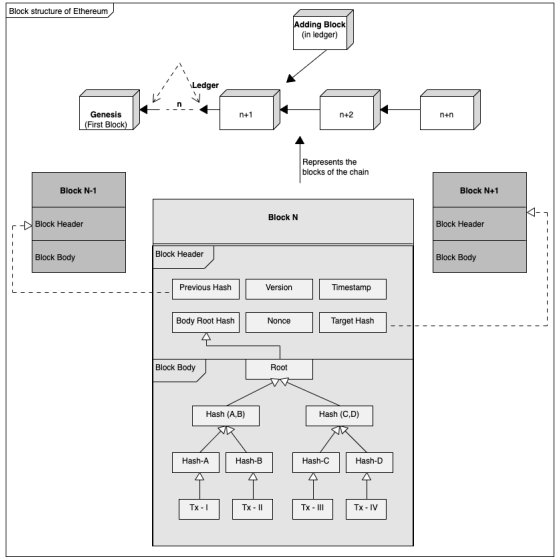

Fig. 2. Ethereum's Block structure and Merkle Root [18]

Fig. 2 also illustrates the block header containing the block version that validate block rules; the hash of the "previous block" represents the value of the previous block and the "timestamp" represents the creation time of the current block. "Body Root Hash" is the root value of the Merkle tree created by transactions in the body of the block, and whereas the "Target Hash" is the hash value threshold of a new valid block. On the other hand, "Block Body" contains verified transactions and all valid transactions are stored in a "Merkle Tree". When the network is created, the "Genesis Block" is automatically assigned with hash default values, and further blocks are added to the ledger after the genesis block. [24]. In addition, the "Ethereum State" is one of the most important aspects of the network consists of accounts, which have a 20-byte address and state transitions that are assigned to a single account [25]. The "World State" is liable for mapping addresses to account states, whereas the Ethereum network's "Consensus" is based on a modified version of the Greedy Heaviest Observed Subtree (GHOST) protocol.

### B. Hyperledger Fabric

Hyperledger [26] is a consensus-based distributed peer-to-peer ledger that combines blockchain with a system for "smart contract" application and other assistive technologies that can be utilized to develop a new generation of transactional applications that focus trust, accountability, and transparency at their core, while streamlining business processes and legitimate limitations. Hyperledger promotes a collaborative approach to developing blockchain applications throughout intellectual property rights and the acceptance of essential standards. [27]. It also embraces the full spectrum of use cases that are crucial for enterprise-based systems. On the same venue, healthcare industry must concentrate on the issue of patient privacy and security of medical records and Hyperledger Fabric includes novel security features such as private data collections, which allow only certain authorized users to access certain data [19, 28]. Unlike open permissionless systems that allow anybody to enroll through a trusted Membership Service Provider in order to participate in the network (MSP). MSP is the technique that allows the remainder of the network to trust and recognize an identity without releasing the member's private key [26]. On another venue, the Hyperledger framework allows the execution of up to 3,500 transactions per second while Ethereum can execute 15 transactions only.

Consensus is an essential component of Hyperledger Fabric which characterizes as a distributed procedure in which a network of nodes ensures that transactions are processed in a guaranteed unique order and verifies transaction blocks [29]. It enables to foreordain the varieties of channels, peers and consensus procedures needed for implementing and testing the proposed approach. These features provide another layer of security, ensuring that resources can only be accessed by network members and network transactions. Therefore, the administrator controls who can join the network and what roles they can play, as well as the ability to delete nodes if necessary.

### C. Analogy Between Hyperledger Fabric and Ethereum

The Linux Foundation hosts Hyperledger Fabric, one of the most prominent blockchain frameworks. It provides plug-and-play components, such as consensus and membership services, as the basis for a solution with a modular architecture. Ethereum, on the other hand, is a decentralized platform that allows smart contracts and apps to execute without the risk of downtime, censorship, fraud, or third-party interference [30]. Hyperledger Fabric intends to provide a Business-to-Business (B2B) platform with a modular and extendable architecture that can be used in a variety of industries, ranging from banking and healthcare to supply chains. However, Ethereum is mostly Business-to-Commerce (B2C) that advertises itself as agnostic to any given field of application [31]. An analogical summary is given in Table II-C.

Both Ethereum and Hyperledger come with their unique advantages for different business scenarios and challenges. However, based on our extensive investigation, we conclude that Ethereum serves well for public applications; Hyperledger's capabilities seem more appealing in enterprise-based blockchain development. Therefore, Hyperledger Fabric shall be adopt for healthcare-based applications due to its suitability for managing health records compared to Ethereum. Hyperledger provides highly flexible, scalable and confidential infrastructure solutions with explicit anonymity and transaction privacy [20].

TABLE I
ANALOGICAL ILLUSTRATION BETWEEN HYPERLEDGER FABRIC AND ETHEREUM

| Blockchain | | |
|---|---|---|
| Category | Hyperledger Fabric | Ethereum |
| Purpose and Confidentiality | Hyperledger Fabric is designed for B2B businesses with Confidential transactions | Ethereum is designed for B2C businesses and generalized applications with transparent transaction |
| Modularity and Extendibility | Hyperledger Fabric is a flexible and extendable architecture that may be used in a variety of enterprise settings | For Modularity and Extendibility, different approaches need to be adopted for Ethereum |
| Cost-Effective | Hyperledger Fabric is a native platform that enables the creation of a self-contained private ledger and contract management without the use of fees. | Ethereum based blockchains network require a fee for each transaction |
| Scalability | Hyperledger Fabric allows more scalability when organizations are added or removed from a channel | Ethereum's scalability bottleneck is notable since each node in the network has to process each transaction |
| Plug and Play API | Hyperledger Fabric provides SDKs in Node.js and Go that can be used to interact with the blockchain efficiently | Ethereum's plug-and-play modularity allows you to customize privacy and permissions on a single platform |
| Consensus | Hyperledger Fabric consists of different phases of checking consensus and all peers on the network do not have to come to some agreement before a transaction is successful | Proof-of-work (PoW) is a consensus protocol that allows the Ethereum network's nodes to agree on the state of all information recorded on the Ethereum blockchain |
| Security and Privacy | Fabric is a permission network, all nodes participating in the Modular Membership Provider provides an identity to everyone in the network (MSP) | Although Ethereum employs transparency as part of its security, there are concerns about data vulnerability |
| Transaction Speed | Hyperledger Fabric's transaction speed capacity vary from 3000 transactions per second to 20000 which is impressive | Ethereum network can only support approximately 30 transactions per second which are quite narrow as of today |

*D. Blockchain in Software Engineering*

Blockchain is growing at a staggering rate with intrinsic potential in the domain of software application. However, the ad hoc approach to adopting blockchain can lead to situations and results that will not be better, but worse, and shall bring serious disappointment [32]. For instance, blockchain is high on energy consumption along with scalability, lack of interoperability, stand-alone projects, difficulty integrating with legacy systems, and complexity issues [33]. To overcome such issues, systematic frameworks have been proposed by inheriting the software engineering approach which is a systematic, disciplined, and quantified approach to software development, operation, and maintenance method; that is, the application of engineering in software [34]. These frameworks are using practices of a newly developed approach called Blockchain-Oriented Software (BOS) engineering, which is relatively new [35].

A significant example of inheritance of the concept of software engineering in blockchain reflects in "Smart Contract" where Ethereum is typically written using "Solidity". Solidity is an object-oriented language having data structures, public and private functions, and the ability to inherit notions like events and modifiers from other contracts [?]. M. Marchesi et al. [35] demonstrates an Agile Software Engineering technique for designing Blockchain applications in 3.

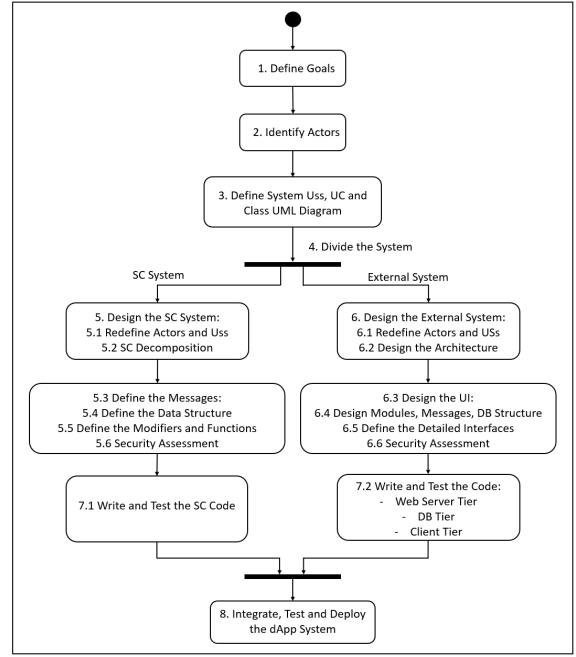

Fig. 3. Proposed Blockchain-Oriented Software development process [35]

Software Engineering provides absolute guidelines at every stage of the software development life cycle, not only for conventional software development but also shall be extended to blockchain-based applications. Challenges in the implementation of such concepts into blockchain shall be broad due to the emergence of new technologies; however, adopting the domain including requirements, process, testing, security, maintenance, configuration management, and verification and validation is a new demand for the blockchain-based software application. Research efforts may be encouraged to provide additional prototypes and proofs-of-concept as a result of research work committed to this issue in order to improve knowledge of blockchain-oriented SE applications [36].

*E. Blockchain in Healthcare*

Blockchain was initially used primarily in the financial industry to allow Bitcoin to function; however, efforts have been made to adapt the technology for a variety of industries, including healthcare, insurance, pharmacy, manufacturing, e-voting, energy, and many more [37]. The healthcare industry is particularly challenge as it has a complex mechanism with various influential stakeholders and a need to disrupt through innovative solutions. Blockchain has applications that can potentially address healthcare issues including public health management, remote monitoring, electronic health records (EHR), medical data management, data security, and drug development. Remarkably, blockchain can mitigate concerns about data ownership and share by allowing patients to keep their data and choose who they share it with [38]. Gaynor, Mark et al. [39] graphically displays several blockchain opportunities for transferring health care data in Fig. 4, where these applications could assist the health care industry to better data exchange across all industry activities, including exchange, storage, and record keeping.

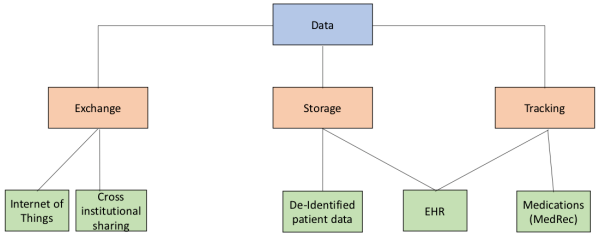

Fig. 4. Data (electronic health record) exchange tree [39]

Blockchain has the ability for addressing significant healthcare concerns while also providing unique chances to leverage the power of other emerging technologies. Despite interoperability challenges including the lack of an existing standard for developing blockchain-based healthcare application, enabling blockchain to solve many complex problems found in today's healthcare industry that shall allow a transformation with the help of researchers and practitioners from different fields towards improving and innovating methods for viewing the health care industry [39, 40].

*F. ONC's Requirements*

The Office of the National Coordinator for Health Information Technology (ONC) introduced the "Cures Act", also referred to as "21st-century cures act" to offer patients and their caregivers seamless and access to confidential information, exchange, and use of electronic health information. Its goal is to promote an ecosystem of new applications to increase innovation and competition in the healthcare industry, giving patients more options [41, 42]. ONC act provides not only the right to electronically access the entirety of their electronic health information and records (EHI, EHR), structured and/or potentially unstructured, at no expense, but also ensure that physicians use technology to offer and exchange electronic health information with patients efficiently [43]. The new provisions certify the third party to access health information in a fair and consistent manner with the permission of the respective stakeholder(s). The applicability of ONC's rules shall be in effect by 2023 as illustrated in Fig. 5; it shall comply by caregivers, IT developers (health), and related stakeholders.

One of the primary purposes of ONC is to create a set of necessary guidelines for health IT developers that need to be followed for establishing secure standards [44]. The new standards and implementation require IT professionals to be certified as health IT developers in order to adequate the newly established technical requirements to provide better connection and user experience of next-generation health modules for providers, as well as facilitate a patient's access to their electronic health record's primary data, IT experts must be qualified as health IT developers.

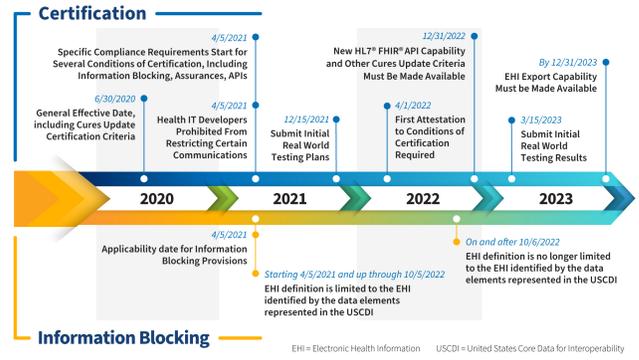

Fig. 5. The Cures Act Final Rule Highlighted Regulatory Dates [39]

## III. RELATED WORK

According to a report of the Health System Tracker [2], the number of aging/older adults is set to augment by 69% (from 56M to 94.7M) in the next 35 years. Hence, improving quality of care and better access to healthcare facilities is important for the society and more so for the elderly population. Especially in pandemic times, when normal lifestyle is disrupted, and the population is expected to stay home, the need for remote patient monitoring has increased and the necessity is larger than ever before. Towards solving the existing problem in healthcare, researchers from different fields proposed different schemes by adopting blockchain technology. After reviewing several studies and systems of blockchain and smart contracts, we identify two frameworks, Ethereum and Hyperledger Fabric. We first carried out a "Search Process" to identify potential studies related to our research using the keywords Blockchain, Telehealth, and Remote Patient Monitoring in order to construct the search string. We used three digital database sources including (i) IEEE Xplore (ii) Research Gate and (iii) Springer Link.

We filtered a specific time period to search studies published between 2016 to 2021 as well as publication topics including health care, and blockchains for IEEE access and computer science discipline (Springer Link). A total of 1,384 studies were found during the initial search. In order to apply the inclusion and exclusion criteria stated in Table, we examine the title, abstract, and conclusion of a research study II - III.

*A. Applications*

- Gem Health Network: In collaboration with Philips Blockchain Lab, a company named "Gem" develops enterprise health care applications networks using blockchain technology [45, 46]. The network includes wellness apps and global patient ID programs that create a healthcare ecosystem using the Ethereum approach. In order to address the trade-off between patient-centric treatment and operational efficiency, the applications would be connected to a universal data infrastructure [47]. As a result, different healthcare operators can access the same information using the Gem Health network that shall include identity schemes, data storage, and smart contracts. In conclusion, the systems solve important operational problems in healthcare industries.

TABLE II
GENERALIZED TABLE FOR SEARCH CRITERIA

| Database | Initial Search | Total Inclusion |
|---|---|---|
| IEEE Xplore | 120 | 10 |
| Research Gate | 25 | 2 |
| Springer Link | 35 | 8 |
| Total | 180 | 20 |

TABLE III
OVERVIEW OF EXCLUSION AND INCLUSION

| Condition of Exclusion and Inclusion | | |
|---|---|---|
| Category | Condition (Inclusion) | Condition (Exclusion) |
| Type | Blockchain, Healthcare, and telehealth based | Based on framework other than said approaches |
| Approach | Studies or Systems that that | that do not discuss or propose approaches |
| Similarity | Research and System provide similar aspects | Studies that does not depict expected aspects |
| Language | Studies that are available in English | Any other languages than English |

- MedRec: MedRec is a decentralized record management system for electronic medical records (EMRs) developed by Ariel Ekblaw and Asaph Azaria using blockchain technology. [48, 49]. The system was developed using Go-Ethereum (Geth) and Solidity; however, it was not built on the live Ethereum network, instead, it creates a small-scale private blockchain with extensive, specific APIs [50]. MedRec makes it simple for patients to access their medical records across providers and treatment venues [49].
- Carechain: A Swedish startup company Carechain, led by IT pioneers Johan Sellström and Stefan Farestam initiated a blockchain-based personal healthcare data management system that intends to focus on protocol level and create a new infrastructure that no one owns, but everyone can control [51]. The Carechain adopted the Ethereum approach with the aim of creating a national blockchain for health data where the system allows individuals ownership and control over their own health information [52]. The system shall assign a universal digital ID owned and controlled by the individual in order to put the individual at the center. The system shall ensure integrated information integrity, built-in policy guarantees with traceability.
- Dovetail: Dovetail is a blockchain-based digital consent application using Hyperledger Fabric approach that allows the sharing of patient data to improve systems related to healthcare, it's products, and services [53]. The system provides a fully audited medical data exchange ledger and harnesses certain specific properties of distributed ledger technology to validate identification, store consent, and establish tamper-proof audit records [19]. Dovetail system stores old medical records and adds new using traditional high-encryption channels with sophisticated data interpolation to transmit patient data and store patient consent for data sharing, and to make sure that that consent is respected by everyone in the process.
- Axuall: Axuall introduces a blockchain-based national digital network powered by the Sorvin Network and Hyperledger Indy to verify identity, credentials, and authenticity in real-time. The system will empower clinicians, medical care, and essential source organizations to share and oversee authenticated credentials, all while adhering to regulatory requirements [54]. The platform automatically aggregates and verifies credentials that can be shared between clinicians and organizations in real-time by ensuring the highest standards of compliance and security. The physicians will be able to provide a fully compliant set of certificates, while healthcare organizations will be able to confirm the legitimacy of a doctor's qualifications [55].
- MedHypChain: MedHypChain is a privacy-preserving, interoperable, patient-centric, hyper-ledger-based medical and data sharing application, that uses an Identity-based broadcast group signcryption mechanism to secure each transaction [56]. The system allows secure implementation of patient-centered interoperability (PCI) data exchange between the patient and medical server remote diagnosis of a patient, interoperability, and malicious participant tracing. The technology achieves great secrecy, anonymity, traceability, and unforgeability, according to the demonstration.

## IV. SYSTEM OVERVIEW

We propose research into the establishment of a decentralized, peer-to-peer network of participants wherein transactions are recorded on a shared distributed ledger for the purpose of patient data management and secure accessibility. Participants in the network would govern and agree by consensus on the updates to the records in the ledger. Also, every record would contain a timestamp and unique cryptographic signature. As a result, the ledger becomes an auditable, immutable record of all network transactions to ensure data security and integrity, while allowing the choice for patients and providers to access the data anytime and anywhere.

## A. System Architecture

Our primary goal is to demonstrate the structure of both blockchain-based frameworks for telehealth and healthcare based application. As part of our scratch, we initially develop a demo prototype to furnish a clear visualization of the proposed blockchain-oriented healthcare application. We adopt Ethereum framework towards a better understanding of how the system mechanism shall function within a Ethereum-based blockchain environment. The developed prototype comprises of (i) a secure Application Programming Interface (API) that meets with the criteria of Office of the National Coordinator for Health Information Technology's (ONC) requirements for facilitating easy access, sharing, and use of patient Electronic Health Records, and (ii) Ethereum based data repository. For demonstration purposes, we focus on Electronic Health Record (EMR) that shall be created in a secure data repository to enable secure upload, store, analyze, retrieve, and transmit patient data according to the patient's instructions, or distribute it when and where it is needed.

derstanding of the design of the proposed system where we present the overall flow of the application consists of four entities. The data shall be collected from the residents either remotely or home healthcare and shall be stored in a decentralized blockchain database. Each data shall be linked with a unique and unchangeable hash and timestamp and shall allow the assigned stakeholders in the retrieval and transmission of stored data using the designed API. An overview of the proposed system illustrates on Fig. 7 while Fig. 8 shows an overview of the interfaces to be provided.

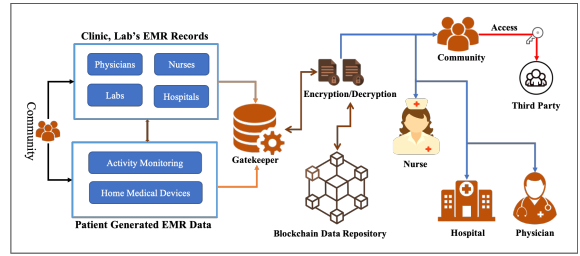

Fig. 7. A low-level architecture of proposed application

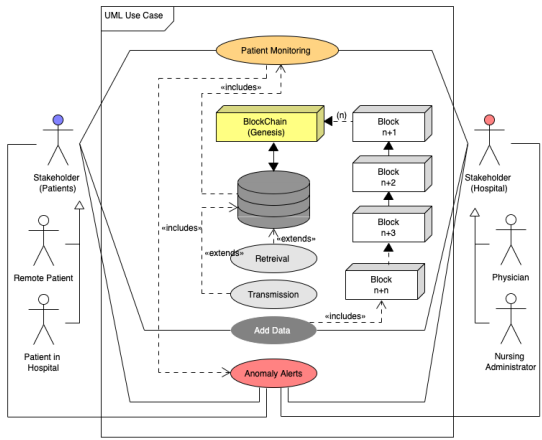

Fig. 6. UML Use Case for Blockchain-Based Health Application

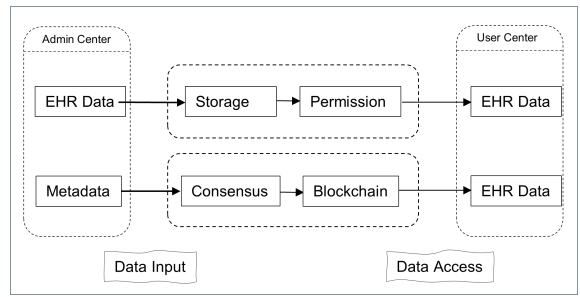

Fig. 8. An overview of metadata

We followed a variety of guidelines from different organizations and approaches– ONC and the concept of software engineering is included. In accordance to SE, we analyze the existing problem, elicit the requirements and identify the research questions by adopting the most suitable standards Agile methodology in blockchain. We develop an UML use case (80%) to scratch down the sequence of actions towards identifying objects and constituting a complete task or transaction in proposed application during requirements analysis or creation process for current research. The use case initializes as a way to capture the main functional goals, and the motivation of laying the architectural foundation for a system, as a result, facilitating requirement coverage [57]. We display four types of stakeholders including (i) Hospital and (ii) Patients in the use case depicted in Fig. 6. The use case also illustrates scores of functions; for instance, patient monitoring and anomaly alert which shall extend in continuous research.

We carried out a low-level architecture for a better un-

Users engage with the proposed platform using a simple online interface in which the metadata has been segregated yet tied together by a unique identifier, resulting in a metadata duple being considered an experiment with a name and an identity. The system will validate metadata by the consensus algorithm that operates among peers within the consortium. Once the metadata is verified by consensus, the metadata is merged into one block and attached to the screwless ledger, which consists of block-shaped metadata entries.. The metadata shall be stored as an object with a timestamp. The peers within a consortium shall allow to search and retrieve the metadata and only be allowed to access that user has permissions. Each block of the proposed system is chained together in an append-only structure using a cryptographic hash function. As a result, altering and erasing recently affirmed data is impossible resulting in new data being appended in the form of additional blocks chained with earlier blocks. Meaning that, updating data of one of the transacted blocks shall generate a different hash value and different link relation towards accomplishing immutability and security.

The proposed system consists of two major modules, organizational and patients. The organizational section enables access to the permissioned stakeholders for adding, updating, retrieving, and monitoring EHR data information while second modules allow patient and assigned representative to access and update certain EHR records. The system was initially developed as a web application that may be extended to mobile applications in future research.

*B. System Entities*

- Blockchain Network: Blockchain network is one of the primary components of the proposed framework intended to store Electronic Health Records (EHR) in a secure decentralized location. Stored EHR data with timestamp and hash shall be generated in the network. The network consists of two types of primary stakeholders responsible to add, update, retrieve data in a blockchain network.
- Cloud Database: We utilize a cloud-based platform named Etherscan, which is a decentralized smart contracts platform suitable for Ethereum. Etherscan allows the users to look up, confirm and validate transaction histories including token transfer and contract execution on the Ethereum decentralized smart contracts platform. Stored data on cloud can be remotely accessed by permissioned stakeholders from anywhere using internet.
- Healthcare Stakeholders: Blockchain can facilitate healthcare stakeholders including physicians, nurses, and patients in different forms. The proposed system enables said authorized stakeholders to access the network that can help to reduce the complexity and security issues.

*C. System Implementation*

We adopt Etherscan which is a block explorer for the proposed platform for discovering, verifying, and approving transactions that occur on the Ethereum blockchain. We utilize the API service of Etherscan for developing decentralised network. Etherscan store data as hash (TxHash) into the block along with the timestamp of real-time based confirmed transaction and its fees.

*D. System Evaluation*

After contriving an initial prototype, we generally validate through experimental simulation approaches. The test environment, observation points, transaction characteristics, workloads, and network size are all important factors to consider when creating a blockchain evaluation. These features should be included in the evaluation findings since revealing them makes it easier to evaluate performance across platforms.

In Fig. 9, an API is depicted where stakeholders can create new patients profile by filling the preset form. Once submitted, the EMR data shall transact to the blockchain storage, details show in Fig. 11. Each EMR data consists of a unique hash and timestamp and the stored data is accessible using the authorized API, in Fig. 10.

Fig. 9. Depicts a form to create patient data

Fig. 10. An interface for accessing stored data using API

Fig. 11. Stored data saved in Ethereum-Based Blockchain Network

## V. DISCUSSION

The outbreak reveals the dire need to invest and improve health infrastructure to better monitor and address the health records of patients. We can see a tendency of fluctuation between the centralization and subsequent decentralization of computing power, storage, infrastructure, protocols, and code if we look back over the last half-century of computer technologies and architectures. Muneeb Ali asserts that We are currently witnessing the shift from centralized computing, storage, and processing to decentralized architectures and systems, which allow us to provide end-users explicit ownership over digital assets while eliminating the need to trust third-party servers and infrastructure. [58]. The frameworks offer different aspects and methodology for every application and choosing the right framework shall depend on requirements specification. Unlike many other applications, healthcare systems' requirements are specific and the primary fundamental is to secure the EMR records in order to prevent manipulation; therefore, the application must embody a rich set of privacy features [59]. The demonstration indicates a positive result in Ethereum based healthcare and telehealth data management. Future research shall continue to facilitate and evaluate Hyperledger Fabric within practical boundaries.

Blockchain technology in software engineering is an emerging field and inheriting the concept of SE into Blockchain is indeed as of today's demand. In this study, we elaborate on some basic inter alia interaction between the aforementioned domain. The perusal indicates the importance of SE in the development of emerging blockchain technology in order to ensure proper guidelines, overcome conventional and security challenges along with the systematic framework for futuristic systems. Such illustrative study encourages researchers to contribute to enhancing the inheritance of software engineering into Blockchain.

## VI. Conclusion

Blockchain is emerged to solve issues people are going through for conventional databases and related existing problems. In this study, we successfully discuss the progress of Blockchain along with an overview of the aspects of Software Engineering. An Ethereum-based system is introduced that has competency in storing Electronic Health Data within a secure and immutable blockchain network. System demonstration indicates that the prototype allows the permissioned stakeholders to add, update, and retrieve EHR data on a RESTful API environment. We recognize possible directions for future research, thus, the concerns revealed in our analysis, we will outline a framework for further empirical research that will be done extensively.

## VII. Acknowledgement

This work was supported in part by research computing resources and technical expertise via a partnership between Kennesaw State University's Office of the Vice President for Research and the Office of the CIO and Vice President for Information Technology [60].